\documentclass[twocolumn,epjc3]{svjour3mod}

\usepackage{amsmath,amssymb,amsfonts,dcolumn,mathcomp,slashed,dsfont}
\usepackage{graphicx}
\usepackage{bm}
\usepackage{xcolor}
\usepackage{csquotes}
\usepackage{units}
\usepackage{flushend}
\usepackage{cite}
\usepackage[colorlinks,citecolor=blue,urlcolor=blue,linkcolor=blue]{hyperref}
\usepackage{ragged2e}

\apptocmd\thebibliography\justifying{}{}

\newcommand{\M}{\mathcal{M}}
\newcommand{\T}{\mathcal{T}}
\newcommand{\F}{\mathcal{F}}
\newcommand{\BR}{\mathcal{B}}
\newcommand{\Lagr}{\mathcal{L}}
\newcommand{\Fpi}{F_\pi}
\newcommand{\Order}{\mathcal{O}}
\newcommand{\disc}{{\rm disc}\,}
\newcommand{\dd}{{\rm d}}
\newcommand{\mpi}{M_{\pi}}
\newcommand{\mpn}{M_{\pi^0}}
\newcommand{\mpc}{M_{\pi^{\pm}}}
\newcommand{\mV}{M_V}
\newcommand{\mw}{M_{\omega}}
\newcommand{\mr}{M_{\rho}}
\newcommand{\Gr}{\Gamma_{\rho}}
\newcommand{\nnnl}{\nonumber\\}
\renewcommand{\Re}{{\rm Re}\,}
\renewcommand{\Im}{{\rm Im}\,}
\newcommand{\wtopi}{\omega\to3\pi}
\newcommand{\mwphys}{\bar{M}_{\omega}}
\newcommand{\mpiphys}{\bar{M}_{\pi}}
\newcommand{\ie}{\textit{i.e.}}
\newcommand{\eg}{\textit{e.g.}}
\newcommand{\cf}{\textit{cf.}}

\newcommand{\GeV}{\,\text{GeV}}
\newcommand{\MeV}{\,\text{MeV}}
\newcommand{\bsp}{\begin{sloppypar}}
\newcommand{\esp}{\end{sloppypar}}

\def\XXint#1#2#3{{\setbox0=\hbox{$#1{#2#3}{\int}$}
     \vcenter{\hbox{$#2#3$}}\kern-0.5\wd0}}

\begin{document}

\title{\boldmath Quark-mass dependence in $\omega\to3\pi$ decays}

\author{Maximilian Dax\thanksref{addr1}
\and
Tobias Isken\thanksref{addr1,addr2,e2} 
\and
Bastian Kubis\thanksref{addr1,addr2}}

\thankstext{e2}{e-mail: isken@hiskp.uni-bonn.de}

\institute{Helmholtz-Institut f\"ur Strahlen- und Kernphysik (Theorie),
   Nussallee 14--16,    
   Universit\"at Bonn, 
   53115 Bonn, Germany\label{addr1}
\and
   Bethe Center for Theoretical Physics,
   Universit\"at Bonn, 
   53115 Bonn, Germany\label{addr2}
}

\date{}

\maketitle

\begin{abstract}
\bsp
We study the quark-mass dependence of $\omega\to3\pi$ decays, based on a dispersion-theoretical framework. 
We rely on the quark-mass-dependent scattering phase shift for the pion--pion $P$-wave extracted from unitarized chiral perturbation theory. 
The dispersive representation then takes into account the final-state rescattering among all three pions.
The described formalism may be used as an extrapolation tool for lattice QCD
calculations of three-pion decays,
for which $\omega\to3\pi$ can serve as a paradigm case.
\esp
\end{abstract}

%-----------------------------------------------------------------------------------------
\section{Introduction}
%-----------------------------------------------------------------------------------------
\bsp
Despite tremendous progress in simulating Quantum Chromodynamics (QCD) on space-time lattices using physical quark masses, many studies of complicated observables within lattice QCD are still performed with light quarks that are heavier than they are in the real world (see \eg\ Refs.~\cite{Aoki:2016frl,Briceno:2017max} for reviews). To extrapolate such simulations to the physical point, additional theoretical input is required, which should ideally be based on systematically improvable effective field theories.  At low energies, the effective field theory that controls the quark-mass dependence by construction is chiral perturbation theory (ChPT)~\cite{Weinberg:1978kz,Gasser:1983yg,Gasser:1984gg}, which describes the interactions of the pseudo-Goldstone bosons of spontaneous chiral symmetry breaking, the pions (as well as kaons and the $\eta$).  

However, the vast majority of states in QCD are resonances, and to perform \textit{chiral extrapolations} for these is less straightforward.  A popular tool in this regard has been to employ unitarized versions of ChPT, such as the inverse amplitude method (IAM)~\cite{Truong:1991gv,Dobado:1992ha,Dobado:1996ps,GomezNicola:2007qj}: a resummation of higher-order effects obeying $S$-matrix unitarity (\ie, probability conservation) allows one to generate poles on unphysical Riemann sheets in the complex-energy plane, the signatures of resonances.  The IAM can be justified using dispersion theory; scattering amplitudes constructed via the IAM match smoothly on the ChPT expansion at low energies.
In this manner, the properties of elastic resonances such as the $f_0(500)$ and the $\rho(770)$ in pion--pion as well as
the $K_0^*(700)$ and $K^*(892)$ in pion--kaon scattering have been investigated with respect to their
quark-mass dependence~\cite{Hanhart:2008mx,Nebreda:2010wv}.

Nevertheless, by far not all hadronic resonances appear in two-body scattering processes.  The lightest resonance that decays only into a three-body final state (in QCD in the isospin limit) is the $\omega(782)$, with its dominant decay $\wtopi$.  Clearly, the quark-mass dependence of the $\omega$ cannot be assessed within an IAM-type formalism; it could at best be studied within the appropriate partial wave of the $3\pi\to3\pi$ scattering process, and the formalism to study such processes on the lattice is currently under intense investigation~\cite{Polejaeva:2012ut,Briceno:2012rv,Hansen:2014eka,Hansen:2015zga,Hammer:2017uqm,Hammer:2017kms,Mai:2017bge}.
In this article, we suggest an approach to assess the quark-mass dependence of the $\wtopi$ decay amplitude based on dispersion relations.  We employ the so-called Khuri--Treiman equations~\cite{Khuri:1960zz} that require the pion--pion two-body phase shift as input, which we extract from the known quark-mass-dependent IAM partial wave.  While we still need to rely on effective field theory ideas to describe the variation of the $\omega$ \textit{mass} with the quark masses, the dispersive framework allows us to predict its quark-mass-dependent \textit{width}.  The idea to employ dispersion theory to extend the applicability of IAM-generated phase shifts is not new: it has already been applied to describe the pion vector form factor~\cite{Guo:2008nc}, as well as, in a formalism closely related to what we present here, to the reaction $\gamma\pi\to\pi\pi$~\cite{Niehus:2017}.
\esp

The outline of the present article is as follows.  
We recall the description of pion--pion scattering with the IAM formalism in Sect.~\ref{sec:rho}.
The Khuri--Treiman formalism for $\wtopi$ is described in Sect.~\ref{sec:dispframework}.
Supplementary assumptions to describe the quark-mass dependence of the $\omega$ width 
are collected in Sect.~\ref{sec:width}, before we show results in Sect.~\ref{sec:results}.
We summarize our findings in Sect.~\ref{sec:summary}.

%-----------------------------------------------------------------------------------------
\section{\boldmath Pion--pion scattering and the $\rho$ resonance in one-loop unitarized ChPT}\label{sec:rho}
%-----------------------------------------------------------------------------------------
Before starting with the discussion of the quark-mass dependence of the $\omega$ as a three-pion resonance, we briefly summarize the investigation of the quark-mass dependence in $\pi\pi\to\pi\pi$ scattering. The results of this section will be an essential input for the study of $\omega\to3\pi$. We follow the formalism introduced in Refs.~\cite{Hanhart:2008mx,Pelaez:2010fj}. For the later purpose our investigation will focus on the $\rho(770)$ resonance, which appears as a pole in the $P$-wave scattering amplitude. We will treat the $\pi\pi$ rescattering as elastic in this whole section.

The partial-wave decomposition of the $\pi\pi\to\pi\pi$ scattering amplitude $\T_{I}$ of definite isospin $I$ is defined by
\begin{equation}
\T_{I}(s,z_{s})=32\pi\sum_{\ell=0}^{\infty}(2\ell+1)P_{\ell}(z_{s})\,t_{\ell}^{I}(s)\,,
\label{eq_pipi_T}
\end{equation}
with the partial-wave amplitude $t_{\ell}^{I}$ of angular momentum $\ell$, the Legendre polynomial $P_{\ell}$, and the $s$-channel scattering angle $z_{s}=\cos\theta_{s}$. Below any inelastic thresholds the partial-wave amplitude is given in terms of the scattering phase shift $\delta_{\ell}^{I}$ only,
\begin{equation}
t_{\ell}^{I}(s) = \frac{\sin\delta_{\ell}^{I}(s)\,e^{i\delta_{\ell}^{I}(s)}}{\sigma(s)}\,,
\label{eq:amplitude_phase_shift}
\end{equation}
where $\sigma(s)=\sqrt{1-4\mpi^2/s}$. Since we are interested in the $P$-wave only ($I=\ell=1$) we will drop the labels for simplicity from now on.

In ChPT the pion mass is given in terms of the light quark masses as an expansion $\mpi^2 = 2B\hat{m}+\Order(m_{q}^{2})$, where the leading term is known as the Gell-Mann--Oakes--Renner (GMOR) relation~\cite{GellMann:1968rz}, with the two light quark masses combined in $\hat{m}=\tfrac{1}{2}(m_{u}+m_{d})$. The constant $B$ is related to the scalar quark condensate in the chiral limit, which measures the strength of spontaneous symmetry breakdown in QCD. We will work in the isospin limit, meaning that $m_{u} = m_{d}$ and $\mpc=\mpn\equiv\mpi$. A brief discussion of the isospin-breaking effects in $\wtopi$ can be found in Sect.~\ref{sec:omegaisospinbreaking}. The GMOR relation implies that studying the quark-mass dependence is equivalent to an investigation of the pion-mass dependence. Hence from now on we will refer to the pion-mass dependence instead. Since we are interested in pion-mass-dependent quantities, it turns out to be useful to define parameters at the physical point as \eg\ $\bar{M}_{\pi}\equiv\mpi^{\text{phys}}$.

In the ChPT power counting the $P$-wave amplitude for $\pi\pi\rightarrow\pi\pi$ scattering up to next-to-leading order can be written as
\begin{equation}
\label{tChPT1}
t_{\rm ChPT}(s)=t_2(s)+t_4(s)+\Order(p^6)\,,
\end{equation}
where $t_{i}$ denotes the contribution of chiral order $p^i$. The $P$-wave projection of the scattering amplitude $\T_{I}$ given in Ref.~\cite{Gasser:1983yg} yields
\begin{align}
\label{t2t4}
t_2(s)&=\frac{s\sigma^2}{96\pi F^2}\,, \notag\\
t_4(s)&=\frac{t_{2}(s)}{48\pi^2F^2}\bigg[s\left(\bar{l}+\frac{1}{3}\right)-\frac{15}{2}\mpi^2\notag\\
&\qquad-\frac{\mpi^4}{2s}\Big(41-2L_\sigma\big(73-25\sigma^2\big)\notag\\
&\qquad\quad +3L_\sigma^2\big(5-32\sigma^2+3\sigma^4\big) \Big)\bigg]+i\sigma\,t_2(s)^2\,,
\end{align}
where we made use of the abbreviation
\begin{equation}
\label{Abbrev_Lsig_sig}
L_\sigma=\frac{1}{\sigma^2}\left(\frac{1}{2\sigma}\log\frac{1+\sigma}{1-\sigma}-1\right)\,.
\end{equation}
The value for pion decay constant in the chiral limit $F$ is taken from the ratio $\Fpi/F=1.064(7)$~\cite{Bazavov:2010hj,Beane:2011zm,Borsanyi:2012zv,Durr:2013goa,Blum:2014tka,Aoki:2016frl}, where $\Fpi = 92.28(9)\MeV$~\cite{Tanabashi:2018} is the pion decay constant at the physical point. For our purpose it is beneficial to work with $F$ instead of $\Fpi$, since $F$ is independent of $\mpi$.  We treat the combination of low-energy constants (LECs) $\bar{l}=\bar{l}_2-\bar{l}_1$, which occurs in the ChPT expression at next-to-leading order, as a free parameter that will be fixed in the following. Note that $\bar{l}$ is also independent of $M_{\pi}$, since the individual mass dependences of $\bar{l}_{1}$ and $\bar{l}_{2}$ cancel~\cite{Gasser:1983yg}.

This amplitude however cannot capture the effects of the $\rho$ resonance, which we expect to be the dominant effect in the $P$-wave above the threshold region, since unitarity is only fulfilled perturbatively ($\Im t_{4} = \sigma |t_{2}|^2$). Furthermore $t_{2}$ and $t_{4}$ are polynomials in $s$ (up to cuts encoded in the $\sigma$ dependence), thus the analytic structure of the standard ChPT expression $t_{\text{ChPT}}$ does not allow for any poles on the second Riemann sheet.

In order to include the $\rho$ resonance into our amplitude we will use the IAM. This method allows us to construct an amplitude that fulfills unitarity exactly. Up to next-to-leading order the IAM yields
\begin{equation}
\label{tIAM1}
t_{\text{IAM}}(s)=\frac{t_2(s)^2}{t_2(s)-t_4(s)}\,,
\end{equation}
which is equivalent to $t_{\rm ChPT}$ up to corrections of $\Order(p^6)$. Note that  crossing symmetry is now only fulfilled perturbatively.

%-----------------------------------------------------------------------------------------
\subsection{\boldmath Pole position and residue}\label{sec:rhopole}
%-----------------------------------------------------------------------------------------
The characteristic properties of the $\rho$ resonance are encoded in the pole position and residue of the amplitude on the second Riemann sheet. By analytic continuation the amplitude on the second sheet can be expressed in terms of the amplitude on the first sheet~\cite{Gribov:2009zz}. We employ the specific representation~\cite{Moussallam:2011zg}
\begin{align}
t^{\text{II}}(s) &=\frac{t^\text{I}(s)}{1-2\hat{\sigma}(s)\,t^{\text{I}}(s)}\,, \notag\\
\hat{\sigma}(s) &= \sqrt{\frac{4\mpi^2}{s}-1} \,, \quad \hat{\sigma}(s\pm i\epsilon) = \mp i \sigma(s) \,,
\label{eq:amp_2nd_sheet}
\end{align}
where $t^{\text{I}}$ and $t^{\text{II}}$ denote the amplitudes on the first and second sheets, respectively. Thus the pole position of the amplitude is determined by 
\begin{equation}
1-2\hat{\sigma}(s_{\text{pole}})\,t^{\text{I}}(s_{\text{pole}})=0\,,
\label{eq:s_pole_def}
\end{equation}
where $s_{\text{pole}}$ corresponds to
\begin{equation}
\label{spole_M_Gamma_rho}
\sqrt{s_{\text{pole}}} = \mr-\frac{i}{2}\Gr\,.
\end{equation}
This allows us to identify the mass $\mr$, as well as the decay width $\Gr$ of the $\rho$ resonance (that is assumed to be a purely elastic resonance with a single decay channel $\pi\pi$). The location of the $\rho$ pole at the physical pion mass stemming from two studies of $\pi\pi$ scattering with Roy-type dispersion relations~\cite{Colangelo:2001df,GarciaMartin:2011jx} is used to constrain the up to now undetermined LEC. This is done in the following way: we minimize the distance of the $\rho$ pole position at the physical point of the IAM amplitude with respect to the most precise extraction of the pole position from the GKPY analysis\footnote{Note that all three determinations of the $\rho$ pole position in Refs.~\cite{Colangelo:2001df,GarciaMartin:2011jx} lead to compatible results for $\bar{l}$.} of Ref.~\cite{GarciaMartin:2011jx}. This yields $\bar{l}=5.73(8)$.

The pole position of the IAM at the physical point is then given by
\begin{align}
\sqrt{s_{\text{pole}}} &= 0.7620(15)\GeV-i0.0778(11)\GeV\,,
\end{align}
which is in good agreement with Refs.~\cite{Hanhart:2008mx,Pelaez:2010fj} and the real part of the pole position coincides with Refs.~\cite{Colangelo:2001df,GarciaMartin:2011jx} within the error bars. Nevertheless we observe a tension in the imaginary part, which is $\sim4\,\MeV$ ($\sim2-3$ standard deviations) larger than the imaginary parts from the Roy analyses. Thus here we reach the limits of the one-loop IAM description with only one free parameter; for more elaborate studies with an $\Order(p^6)$ IAM amplitude containing several LECs, see Refs.~\cite{Pelaez:2010fj,Niehus:2018}. However, for the purposes of our study of $\wtopi$, we consider the one-loop IAM a sufficiently reasonable description of $\pi\pi$ scattering. 

\begin{figure}
	\centering
	\large
	\scalebox{0.68}{\input{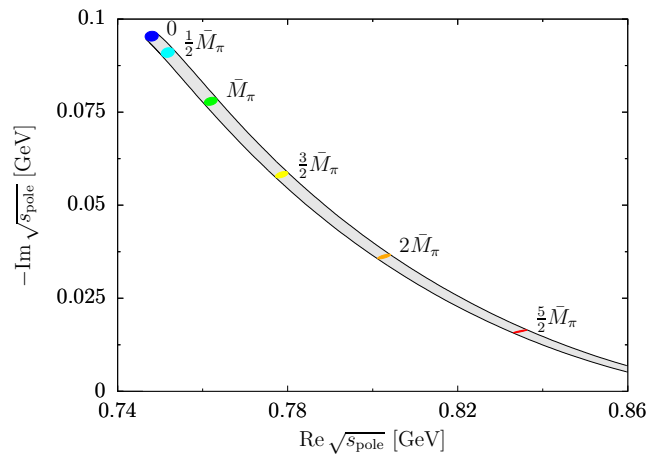}}
	\caption{Trajectory of the $\rho$ pole position of $t^{\text{II}}$ in the complex $s$ plane as given in Eq.~\eqref{eq:amp_2nd_sheet} for different pion masses. The gray error band is generated by the uncertainty of $F$ and $\bar{l}$. The colored ellipses mark the one-$\sigma$ uncertainty regions of the pole position for the respective pion mass.}
	\label{Fig:poleTrajectory}
\end{figure}

Since the amplitude is not limited to the physical value of the pion mass, we are able to calculate $\mr$ and $\Gr$ as a function of $\mpi$. The trajectory of the pole position on the second Riemann sheet is displayed in Fig.~\ref{Fig:poleTrajectory}. As expected from its quark content, the mass of the $\rho$ increases if the pion becomes heavier. This behavior can be described to good approximation as a linear function in $\mpi^2$ given by
\begin{equation}
\mr(\mpi^2) = \mr(0) +a \, \mpi^2\,,
\end{equation}
where $\mr(0)$ and $a$ can be matched to the pion-mass-dependent pole trajectory extracted from Eq.~\eqref{eq:s_pole_def}, which yields
\begin{align}
\mr(0) = 0.7480(16)\GeV\,,\quad a = 0.719(9)\GeV^{-1}\,. \label{eq:Mrho-Mpi-num}
\end{align}
Similar observations have been made by investigating chiral symmetry constraints~\cite{Bruns:2017gix}.

\bsp
The available phase space for the decay decreases with growing pion mass, since the $\rho$ mass increases much more slowly than the pion mass~\cite{Hanhart:2008mx}. Thus the width of the $\rho$ becomes smaller for larger values of $\mpi$. The coupling $g_{\rho\pi\pi}$ of the $\rho$ to the $\pi\pi$ system is defined via the residue
\begin{equation}
g_{\rho\pi\pi}^{2} = -48\pi\lim_{s\to s_{\text{pole}}} \frac{s-s_{\text{pole}}}{s-4\mpi^2}\,t^{\text{II}}(s)\,,
\label{eq:g_rhopipi_res}
\end{equation}
where the normalization factors are chosen such that it coincides with the naive expression
\begin{equation}
\Gr = \frac{|g_{\rho\pi\pi}|^2}{48\pi\mr^2}\big(\mr^2-4\mpi^2\big)^{3/2} \,,
\label{eq:g_rhopipi_VMD}
\end{equation}
as obtained from a Lagrangian-based narrow-width approximation or a vector-meson-dominance (VMD) model. Equation~\eqref{eq:g_rhopipi_res} yields a numerical value of $|g_{\rho\pi\pi}|=6.12(4)$, which is in fair agreement with other determinations~\cite{Pelaez:2010fj,GarciaMartin:2011jx}. Note that the coupling $g_{\rho\pi\pi}$ extracted from the IAM is pion-mass independent to very good approximation. Thus the pion-mass dependence of $\Gr$ is driven by the phase space factor only, see Eq.~\eqref{eq:g_rhopipi_VMD}, as confirmed by lattice QCD calculations~\cite{Alexandrou:2017mpi}.
\esp

%-----------------------------------------------------------------------------------------
\subsection{Scattering phase shift}\label{sec:phaseshift}
\begin{figure}
	\centering
	\large
	\scalebox{0.68}{\input{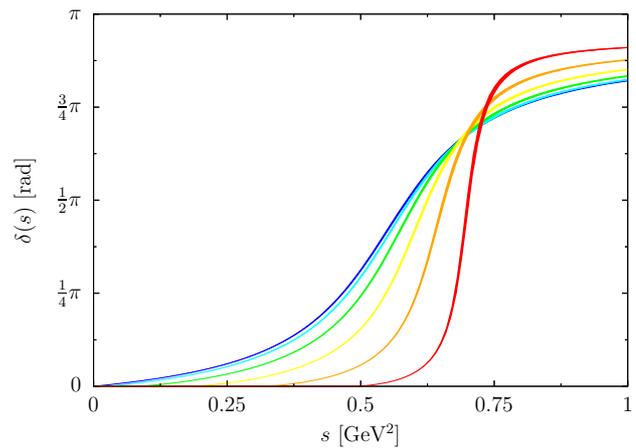}}
	\caption{IAM amplitude $\pi\pi$ $P$-wave scattering phase shift for different values of the pion mass: 0 (blue), $\tfrac{1}{2}\bar{M}_{\pi}$ (cyan), $\bar{M}_{\pi}$ (green), $\tfrac{3}{2}\bar{M}_{\pi}$ (yellow), $2\bar{M}_{\pi}$ (orange), and $\tfrac{5}{2}\bar{M}_{\pi}$ (red). The error bands result from the uncertainty of $F$ and $\bar{l}$. The color coding is identical to Fig.~\ref{Fig:poleTrajectory}.}
	\label{Fig:IAMphase}
\end{figure}

%-----------------------------------------------------------------------------------------
The $I=1$ $\pi\pi$ system is one of the most widely-studied resonant scattering phase shifts in lattice QCD~\cite{Aoki:2007rd,Feng:2010es,Aoki:2011yj,Lang:2011mn,Dudek:2012xn,Pelissier:2012pi,Bali:2015gji,Wilson:2015dqa,Bulava:2016mks,Alexandrou:2017mpi,Hu:2017wli,Andersen:2018mau}. As defined in Eq.~\eqref{eq:amplitude_phase_shift} the scattering phase shift of the IAM amplitude can be extracted via $\delta(s)\equiv\arg t(s)$. The results for different values of the pion mass are shown in Fig.~\ref{Fig:IAMphase}. As expected, the slope of the phase shift becomes steeper for heavier pions, while the whole curve moves to the right (decreasing width and increasing mass of the $\rho$). This behavior is also observed by various lattice QCD calculations carried out at different pion masses~\cite{Aoki:2007rd,Feng:2010es,Aoki:2011yj,Lang:2011mn,Dudek:2012xn,Pelissier:2012pi,Bali:2015gji,Wilson:2015dqa,Bulava:2016mks,Alexandrou:2017mpi,Hu:2017wli,Andersen:2018mau}. At the physical pion mass the phase shift is in perfect agreement with the Roy analyses of Refs.~\cite{Caprini:2011ky,GarciaMartin:2011cn} in the low-energy regime up to $\sqrt{s}\sim0.8\GeV$.

Above this energy the Roy solutions are typically continued to an asymptotic value of $\pi$. The IAM amplitude on the other hand behaves like 
\begin{equation}
\lim_{s\to\infty}t_{\text{IAM}}(s) = -\frac{3\pi}{6\bar{l}+2+3\pi i} 
\end{equation}
in the high-energy limit, which depends only on the value of the LEC. Thus the phase of the IAM amplitude will not reach $\pi$ for all reasonable values of $\bar{l}$. 
Nevertheless the phase shift gives a reasonable parametrization up to the $\rho$ resonance region and thus will be used as a key ingredient for the dispersive representation of the $\omega\to3\pi$ amplitude, as described in Sect.~\ref{sec:dispframework}. In order to test the effect of the discrepancy of the IAM compared to the Roy solution phase shifts, we will also use the parametrization
\begin{equation}
\delta(s) = \pi-\frac{\delta_{a}}{\delta_{b}+s/\Lambda^{2}}
\end{equation}
to account for the correct asymptotic behavior. The parameters $\delta_{a}$ and $\delta_{b}$ are fixed by ensuring continuity of the phase shift and its derivative at some high-energy scale $\Lambda^2\sim1.2\GeV^2$, at which we switch from the IAM to the asymptotic phase shift. Note that this high-energy continuation has no relevant influence on the $\wtopi$ amplitude, since its effects can be fully absorbed by adjusting the unknown normalization of the amplitude, see Eq.~\eqref{eq_basis_functions} below. Similarly, any hypothetical, sub-leading, pion-mass dependence in $\Lambda^2$ would be far too small to be of relevance.

%-----------------------------------------------------------------------------------------
\section{Formalism and dispersive representation of the $\wtopi$ amplitude}
\label{sec:dispframework}
%-----------------------------------------------------------------------------------------
The transition amplitude for the $\omega\to3\pi$ decay is defined as
\begin{align}
&\langle \pi^{+}(p_{1})\pi^{-}(p_{2})\pi^{0}(p_{3})|T|\omega(P)\rangle \notag\\
&\hspace{1cm}= (2\pi)^4\delta^{(4)}(P-p_{1}-p_{2}-p_{3})\,\M(s,t,u)\,.
\end{align}
In our convention the Mandelstam variables are defined according to
\begin{equation}
s=(P-p_{3})^2\,,\quad t=(P-p_{1})^2\,,\quad u=(P-p_{2})^2\,,
\end{equation}
which fulfill the relation
\begin{equation}
s+t+u=\mw^2+3\mpi^2 =: 3s_{0}.
\end{equation}
In the $s$-channel center-of-mass system, $t$ and $u$ can be expressed as
\begin{equation}
t(s,z_{s}) = u(s,-z_{s}) = \frac{1}{2}\big(3s_{0}-s + z_{s}\kappa(s)\big)\,,
\end{equation}
where $z_s=\cos\theta_{s}$ is the scattering angle
\begin{align}
z_{s}=\cos\theta_{s}=\frac{t-u}{\kappa(s)}\,,\quad\kappa(s) = \sigma(s)\,\lambda^{1/2}(\mw^2,\mpi^2,s)\,,
\end{align}
with the K\"all\'en function $\lambda(x,y,z) = x^2+y^2+z^2-2(xy+xz+yz)$. Similar expressions hold for the $t$ and $u$-channels, respectively. The physical thresholds in the three channels are
\begin{equation}
s_{\text{thr}}=t_{\text{thr}}=u_{\text{thr}}=4\mpi^2\,.
\end{equation}

Since the transition $\omega\to3\pi$ is of odd intrinsic parity, the amplitude can be further decomposed into a kinematic prefactor and a scalar function $\F(s,t,u)$ containing the dynamical information,
\begin{equation}
\M(s,t,u) = i\epsilon_{\mu\nu\alpha\beta}\,\epsilon^{\mu}(P)\,p_{1}^{\nu}\,p_{2}^{\alpha}\,p_{3}^{\beta}\,\F(s,t,u)\,.
\label{eq_decopm_M_F}
\end{equation}
Here $\epsilon^{\mu}(P)$ denotes the polarization vector of the $\omega$ meson. The modulus of the amplitude is given by
\begin{align}
|\M(s,t,u)|^2 = \frac{1}{4}\Big[stu-\mpi^2\big(\mw^2-\mpi^2\big)^2\Big]|\F(s,t,u)|^2\,.
\label{eq:kin_prefactor}
\end{align}
The expression in the square brackets is also known as the \textit{Kibble cubic}~\cite{Kibble:1960zz}. 

Due to Bose symmetry only odd partial waves are allowed to contribute to the process. Thus the partial-wave decomposition for the scalar function $\F(s,t,u)$ in the $s$-channel reads
\begin{equation}
\F(s,t,u) = \sum_{\ell~\text{odd}}P'_{\ell}(z_s)\,f_{\ell}(s)\,,
\end{equation}
where $P'_{\ell}(z_s)$ denotes the differentiated Legendre polynomials. A particular partial wave can be projected out by making use of
\begin{equation}
f_{\ell}(s) = \frac{1}{2}\int_{-1}^{1}\dd z_{s}\big[P_{\ell-1}(z_{s})-P_{\ell+1}(z_{s})\big]\F(s,t,u)\,. \label{eq_PW}
\end{equation}
As the available phase space in the $\omega\to3\pi$ decay is rather small, the dominant contribution will come from the $\ell=1$ partial wave (see Ref.~\cite{Niecknig:2012sj} for a discussion of potential $F$-wave contributions). Neglecting discontinuities from $F$- and higher partial waves allows us to decompose the scalar function into a sum of single-variable functions~\cite{Niecknig:2012sj,Hoferichter:2012pm,Danilkin:2014cra,Hoferichter:2014vra,Hoferichter:2017ftn,Hoferichter:2018kwz}
\begin{equation}
\F(s,t,u) = \F(s)+\F(t)+\F(u)\,,
\label{eq_reconstruction_thm}
\end{equation}
where $\F(s)$ possesses only a right-hand cut. This kind of decomposition is known as a \textit{reconstruction theorem}~\cite{Stern:1993rg,Knecht:1995tr,Ananthanarayan:2000cp,Zdrahal:2008bd}. The symmetry in this decomposition reflects the fact that the process is invariant under the exchange of the pions. 
Combining Eqs.~\eqref{eq_PW} and \eqref{eq_reconstruction_thm} leads to
\begin{align}
f_{1}(s) &=\F(s)+\hat{\F}(s) \,, \quad
\hat{\F}(s) = 3\big\langle(1-z_{s}^2)\F\big\rangle(s)\,, \notag\\
\big\langle z_{s}^{n}\F\big\rangle(s) &= \frac{1}{2}\int_{-1}^{1}\dd z_{s}\,z_{s}^{n}\F\big(t(s,z_{s})\big)\,. \label{eq_inhomogeneity}
\end{align}
The right-hand cut of the partial wave $f_1(s)$ is contained in $\F(s)$, while its left-hand cut contributions
reside entirely in the projection of the crossed-channel single-variable functions $\hat{\F}(s)$.

\bsp
The dispersive framework to describe this decay was already used in previous studies of $\omega\to3\pi$~\cite{Niecknig:2012sj,Danilkin:2014cra} as well as the closely related processes $\gamma^{(*)}\to3\pi$~\cite{Hoferichter:2012pm,Hoferichter:2014vra,Hoferichter:2017ftn,Hoferichter:2018dmo,Hoferichter:2018kwz}.  It uses the formalism of the so-called Khuri--Treiman equations~\cite{Khuri:1960zz,Bronzan:1963mby}, which is based on analytic continuation of a crossed scattering amplitude in the decay mass~\cite{Gribov:1962fu}.  Final-state pion--pion rescattering is assumed to be elastic, and can be described in terms of the phase shift only, see Eq.~\eqref{eq:amplitude_phase_shift}.
The unitarity relation for the partial wave $f_1(s)$ is given by\footnote{Here and in the following relations that involve the discontinuity are always meant to be valid along the right-hand cut only, which starts at the two-pion threshold in the respective channel.} 
\begin{equation}
\disc f_1(s) = 2i f_1(s) \sin\delta(s)\,e^{-i\delta(s)}\,,
\label{eq_unitarity_f1}
\end{equation}
where $\delta(s)$ is the $\pi\pi$ $P$-wave phase shift.  Inserting Eq.~\eqref{eq_inhomogeneity} and noting that
$\disc f_{1}(s) = \disc \F(s)$ along the right-hand cut, we conclude
\begin{equation}
\disc\F(s) = 2i\big[\F(s)+\hat{\F}(s)\big]\sin\delta(s)\,e^{-i\delta(s)}\,,
\label{eq_unitarity_F}
\end{equation}
which is an inhomogeneous Omn\`es problem for the single-variable function $\F(s)$
with the inhomogeneity $\hat{\F}(s)$.
Assuming the Froissart--Martin bound~\cite{Froissart:1961ux,Martin:1962rt}, a solution of the unitarity relation~\eqref{eq_unitarity_F} can be written in terms of a single subtraction constant $\alpha$~\cite{Niecknig:2012sj},
\begin{align}
\F(s) &= \alpha\F_{\alpha}(s)\,,\qquad\hat{\F}(s)=\alpha\hat{\F}_{\alpha}(s)\,, \notag\\
\F_{\alpha}(s) &= \Omega(s)\bigg\{1+\frac{s}{\pi}\int_{4\mpi^2}^{\infty}\frac{\dd s'}{s'}\frac{\hat{\F}_{\alpha}(s')\,\sin\delta(s')}{|\Omega(s')|(s'-s)}\bigg\}\,,
\label{eq_basis_functions}
\end{align}
where the Omn\`es function $\Omega(s)$ is given by~\cite{Omnes:1958hv}
\begin{equation}
\Omega(s) = \exp\bigg\{\frac{s}{\pi}\int_{4\mpi^2}^{\infty}\frac{\dd s'}{s'}\frac{\delta(s')}{(s'-s)}\bigg\}\,.
\end{equation}
The \textit{basis function} $\F_{\alpha}(s)$ can be constructed independently of the numerical value of the subtraction constant $\alpha$, which therefore can be determined \textit{a posteriori}.  As $\alpha$ serves as an overall normalization of the amplitude, at physical pion masses it is fixed to the total rate $\Gamma(\omega\to3\pi)$, with the energy dependence of the amplitude or the Dalitz plot distribution then being a theoretical prediction~\cite{Niecknig:2012sj}.
\esp

%-----------------------------------------------------------------------------------------
\section{\boldmath Decay width and $\mpi$ dependencies}\label{sec:width}
%-----------------------------------------------------------------------------------------
The process $\wtopi$ gives the by far dominant contribution to the total $\omega$ decay width, $\BR(\wtopi)=(89.2\pm0.7)\%$. Besides that, the main subleading contributions stem from $\BR(\omega\to\pi\gamma)=(8.4\pm0.2)\%$ (electromagnetic) and $\BR(\omega\to\pi\pi)=(1.5\pm0.1)\%$ (isospin-breaking). Together these contributions account for $>99\%$ of the decay width at the physical point~\cite{Tanabashi:2018}. As long as we restrict ourselves to the isospin limit and strong contributions only, the decay width is fully driven by $\wtopi$.

Thus the decay width $\Gamma(\wtopi) \equiv \Gamma_{\omega}$ is obtained by integrating the squared amplitude over phase space according to
\begin{equation}
\label{Eq_DecWid_LorInvPhaseSpInt}
\Gamma_{\omega}=\frac{1}{256\pi^3\mw^3}\int\dd s\,\dd t\,|\M(s,t,u)|^2\,.
\end{equation}
This expression has several $\mpi$ dependencies besides the explicit ones (integration boundaries and Mandelstam variables), which will be discussed in the following.

%-----------------------------------------------------------------------------------------
\subsection{Pion-mass dependence of the $\omega$ mass}\label{sec:omegamass}
%-----------------------------------------------------------------------------------------
\bsp
In contrast to the case of the $\rho$, for which we can derive the pion-mass dependence of the complete 
pole position in the $\pi\pi\to\pi\pi$ $P$-wave amplitude by means of the IAM, we are not in the position
to do the same for the $\omega$ within some $3\pi\to3\pi$ amplitude of the appropriate quantum numbers.
We will discuss the complicated pion-mass dependence of the \textit{width} of the $\omega$, or the imaginary
part of its pole in the complex plane, which is the 
main focus of this study, in the following; for the pion-mass dependence of its \textit{mass}, the 
corresponding real part, we have to resort to symmetry arguments based on effective Lagrangians.
These will relate $\mw(\mpi^2)$ to $\mr(\mpi^2)$, which we have discussed in Sect.~\ref{sec:rhopole}.
\esp

\bsp
We briefly recapitulate the analysis of the leading symmetry-breaking effects 
in the masses of the vector meson nonet~\cite{Jenkins:1995vb,Bijnens:1996nq,Bijnens:1997ni}.
Here, the vector mesons are treated as static matter fields; the effective Lagrangian is organized
in terms of increasing chiral dimension as well as using the expansion in the inverse number of colors
$1/N_c$.  We neglect isospin breaking and electromagnetic effects~\cite{Bijnens:1996nq}, and ignore
deviations from ideal mixing.  In this approximation, the symmetry-breaking part of the 
effective Lagrangian can be written as
\begin{align}
\Lagr_{\textrm{SB}} &= \frac{\delta}{2} \langle W_\mu^\dagger\rangle \langle W^\mu \rangle 
+ \frac{a}{2}\langle \chi \{W_\mu^\dagger,W^\mu\} \rangle \nnnl
&\quad+ \frac{b}{4} \big( \langle \chi W_\mu^\dagger\rangle \langle W^\mu\rangle + \text{h.c.} \big)
+ \frac{c}{2}\langle \chi \rangle \langle W_\mu^\dagger W^\mu \rangle  \nnnl
&\quad+ \Order(m_q^2,1/N_c^2)\,, \label{eq:Lagr}
\end{align}
where
\begin{equation}
W_\mu = \left( \begin{array}{ccc}
\frac{\rho^0_\mu}{\sqrt{2}} + \frac{\omega_\mu}{\sqrt{2}} & \rho_\mu^+ & K_\mu^{*+} \\
\rho_\mu^- & - \frac{\rho^0_\mu}{\sqrt{2}} + \frac{\omega_\mu}{\sqrt{2}} & K_\mu^{*0} \\
K_\mu^{*-} & \bar K_\mu^{*0} & \phi_\mu
\end{array} \right)
\end{equation}
contains the (nonrelativistic) vector-meson fields, and $\chi = \text{diag}(\mpi^2,\mpi^2,2M_K^2-\mpi^2)$
breaks $SU(3)$ flavor symmetry due to the different quark masses.
Among the terms in Eq.~\eqref{eq:Lagr}, the quark-mass-independent 
operator $\propto \delta$ is $1/N_c$ suppressed and 
breaks nonet symmetry; the term $\propto a$ is chirally suppressed, but the dominant flavor-breaking
term in the large-$N_c$ limit; and the operators $\propto b$ and $\propto c$ are both chirally and 
$1/N_c$ suppressed.  The term $\propto c$ leads to a common shift in all nonet masses and hence
cannot be discerned from the common mass $M_V$ using experimental data only; on account of the fact
that we can show the operator $\propto b$ indeed to be strongly suppressed below, we will neglect 
the former in the following.  
\esp

Equation~\eqref{eq:Lagr} then leads to the vector-meson masses
\begin{align}
\mr &= \mV + a \,\mpi^2\,, \nnnl 
\mw &= \mV + \delta + (a+b)\mpi^2\,, \nnnl
M_\phi &= \mV + \frac{\delta}{2} +  \Big(a+\frac{b}{2}\Big)\big(2M_K^2-\mpi^2\big)\,, \nnnl
M_{K^*} &= \mV + a\, M_K^2\,,
\label{Eq:mV_mpi_dependence}
\end{align}
which allows us to extract the coupling constants according to
\begin{align}
a &= \frac{M_{K^*}-\mr}{M_K^2-\mpi^2}\,,\nnnl
b &= \frac{M_\phi-2M_{K^*}+\frac{3}{2}\mr-\frac{1}{2}\mw}{M_K^2-\mpi^2}\,, \nnnl
\delta &= \mw-\mr - b\,\mpi^2\,.
\end{align}
In particular the value for $b$ depends quite sensitively on the precise values inserted for 
the masses of the broad $\rho$ and $K^*$ resonances; if we employ the real parts of their
pole positions~\cite{GarciaMartin:2011jx,Pelaez:2016klv}, we find
$a = 0.57(1)\GeV^{-1}$, $b = -0.045(20)\GeV^{-1}$,  $\delta = 20(2)\MeV$.  

\bsp
We hence conclude that the formal $1/N_c$ suppression of $b/a$ translates, in fact, into a numerical
suppression by more than an order of magnitude; we will therefore neglect $b$, too.  
Furthermore, we observe that the determination of $a$ based on $SU(3)$ symmetry leads to an estimate
that is about 20\% smaller than the value deduced from the one-loop IAM representation, 
see Eq.~\eqref{eq:Mrho-Mpi-num}.  As we expect that $SU(3)$ breaking effects ought to affect the 
relation between $\rho$ and $\omega$ observables less, we use the arguments above to employ
a pion-mass dependence of $\mw$ that equals the one of $\mr$ up to the constant offset $\delta$,
hence
\begin{align}
\label{Eq_kin_Mw_mp}
\mw(\mpi^2) &=\mwphys+\frac{0.719(9)}{\unit{GeV}}\left(\mpi^2-\mpiphys^2\right) \,, \notag\\
\mw(0) &= 0.7686(20)\GeV \,.
\end{align}

At higher orders in the chiral expansion, Goldstone-boson loops induce nonanalytic dependencies 
of the vector meson masses of the form $\Order(m_q^{3/2})$ and $\Order(m_q^2\log m_q)$, 
which have been studied extensively in the literature~\cite{Jenkins:1995vb,Bijnens:1996nq,Bijnens:1997ni,Bruns:2004tj,Bruns:2013tja,Bavontaweepanya:2018yds}; such terms will obviously break the similarity in 
$\mr(\mpi^2)$ and $\mw(\mpi^2)$ due to the different coupling of $\rho$ and $\omega$ to pions.
We ignore such terms in the present study solely based on the observation in Sect.~\ref{sec:rhopole}
that a linear dependence of $\mr$ on $\mpi^2$ is sufficient to describe the behavior of the $\rho$ pole
of the $\Order(p^4)$ IAM amplitude.
\esp

%-----------------------------------------------------------------------------------------
\subsection{Subtraction constant}
%-----------------------------------------------------------------------------------------
As derived in Sect.~\ref{sec:dispframework}, we require one subtraction constant in the dispersive representation of the $\wtopi$ decay amplitude in order to maintain a convergent integral representation. Since this subtraction constant is not fixed by unitarity (and, for the process at hand, cannot be matched to ChPT as for other processes such as $\gamma\pi\to\pi\pi$~\cite{Hoferichter:2012pm,Niehus:2017}), we need to fix its pion-mass dependence in a different way.

\begin{figure}
\centering
\includegraphics[width=0.45\linewidth]{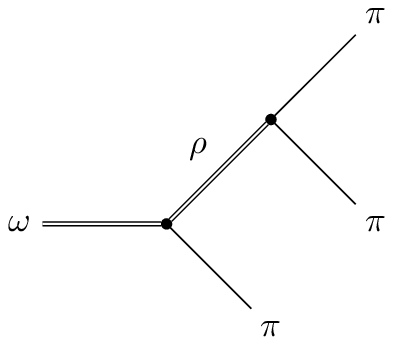}
\caption{VMD tree-level diagram for $\wtopi$ given by a $\omega\to\rho\pi$ and a subsequent $\rho\to\pi\pi$ decay.}
\label{fig:VMD_omega_3pi}
\end{figure}

In Sect.~\ref{sec:rho} we have recounted that the coupling $g_{\rho\pi\pi}$ at the $\rho\to\pi\pi$ vertex is (essentially) pion-mass independent and in good agreement with a narrow-width formula or a VMD model. In an isobar model of subsequent two-body decays, $\omega\to3\pi$ is typically understood in terms of processes $\omega\to\rho\pi$, followed by $\rho\to\pi\pi$ decays, see Fig.~\ref{fig:VMD_omega_3pi} (\cf\ \eg\ Refs.~\cite{GellMann:1962jt,Klingl:1996by,Leupold:2008bp}).  Reducing the dispersive representation Eq.~\eqref{eq_basis_functions} to such a simplified picture, we find the subtraction constant $\alpha$ in one-to-one correspondence with the product of coupling constants $g_{\omega\rho\pi}\times g_{\rho\pi\pi}$.  We therefore conjecture that, by analogy, it is reasonable to assume $g_{\omega\rho\pi}$, and hence $\alpha$, to be also pion-mass independent, and we fix the subtraction constant to the total decay width $\Gamma_\omega$ at the physical point.

%-----------------------------------------------------------------------------------------
\subsection{Isospin-breaking effects}\label{sec:omegaisospinbreaking}
%-----------------------------------------------------------------------------------------
Up to now all calculations have been carried out under the assumption of isospin symmetry ($\mpc=\mpn$). We now want to briefly discuss the influence of isospin breaking effects in $\omega\to3\pi$; a more detailed discussion can be found in Ref.~\cite{Dax:2017}.  It is well known that in the context of precision analyses of $\eta\to3\pi$, in particular comparing $\eta\to\pi^+\pi^-\pi^0$ and $\eta\to3\pi^0$, taking into account the pion-mass difference for the available phase space at least is mandatory~\cite{Gasser:1984pr,Ditsche:2008cq,Schneider:2010hs,Colangelo:2018jxw}. In this section, we therefore only investigate isospin breaking in the kinematical contribution, ignoring all dynamical effects (\ie\ using $|\F(s,t,u)|^2=const.$), expecting this to be the dominant change. 

The pion-mass difference
\begin{equation}
\Delta_{\pi} = \big(\mpc^2-\mpn^2\big) = 1.26116(15)\times10^{-3}\GeV^{2}
\end{equation}
originates from two sources: electromagnetic effects and the light-quark-mass difference $(m_{u}-m_{d})$. 
At leading-order in ChPT (for three flavors), the latter can be evaluated to
\begin{align}
\Delta_{\pi}^{\text{QCD}} = \frac{(m_{u}-m_{d})^2}{8\hat{m}(m_{s}-\hat{m})}\mpi^2 \approx 3 \times10^{-5}\GeV^{2} \,,
\end{align}
hence this effect is very small: the pion-mass difference is dominantly caused by electromagnetism, and we neglect the effect of the difference of the light quark masses completely. Consequently, as long as we neglect higher-order corrections of $\Order(e^2m_q)$, the pion-mass difference $\Delta_{\pi}$ stays constant when varying $\mpi$. This allows us to relate the neutral pion mass to the charged one according to
\begin{equation}
\mpn(\mpc^{2}) = \sqrt{\mpc^2-\Delta_{\pi}}\,.
\label{Eq:mpn_mpc_isobreak}
\end{equation}
Obviously, this relation breaks down at a minimal charged pion mass of $\mpc = \sqrt{\Delta_\pi} \approx 35.5\MeV$.

It turns out that isospin breaking in the kinematical dependence of $\Gamma_{\omega}$ as given in Eq.~\eqref{Eq_DecWid_LorInvPhaseSpInt} when tuning the charged pion mass gives only a tiny correction to $\Gamma_{\omega}$ of less than $2\%$.  Due to the connection of the pion masses in Eq.~\eqref{Eq:mpn_mpc_isobreak} it is easy to see that the effect of isospin breaking becomes smaller when increasing the mass of the pions; on the other hand, since the ratio of the $\omega$ mass to the pion masses becomes larger when approaching the chiral limit, the increasing isospin-breaking effects are lifted by the $\omega$ mass. Thus using the isospin limit is entirely justified for our purposes.

%-----------------------------------------------------------------------------------------
\section{Results}\label{sec:results}
%-----------------------------------------------------------------------------------------
\begin{figure*}
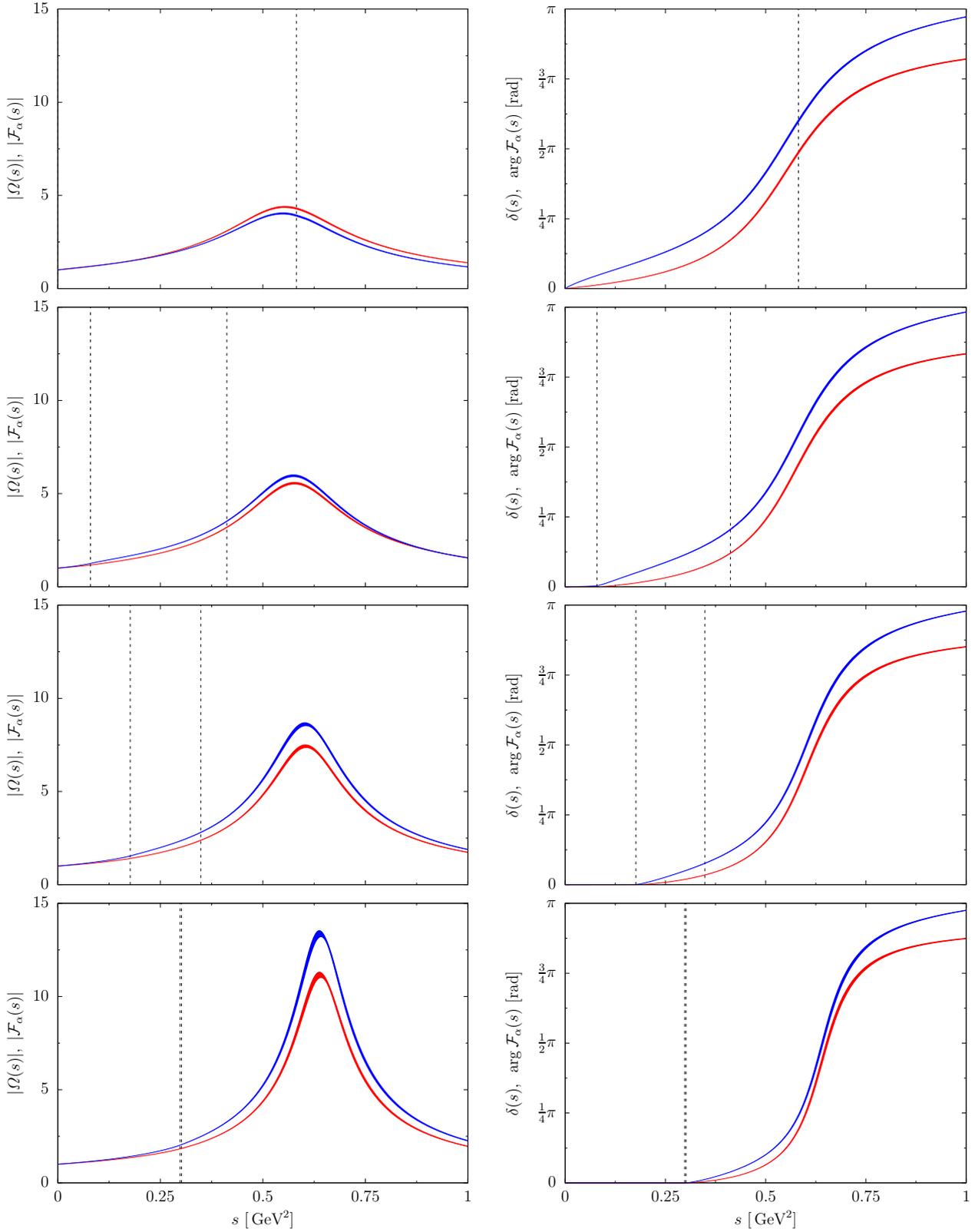

	\large
	\centering
	\scalebox{0.665}{\input{plots/KT_omnes_abs_1.tex} \input{plots/KT_omnes_arg_1.tex}}
	
	\vspace{-0.8cm}
	
	\scalebox{0.665}{\input{plots/KT_omnes_abs_2.tex} \input{plots/KT_omnes_arg_2.tex}}
	
	\vspace{-0.8cm}
	
	\scalebox{0.665}{\input{plots/KT_omnes_abs_3.tex} \input{plots/KT_omnes_arg_3.tex}}
	
	\vspace{-0.8cm}
	
	\scalebox{0.665}{\input{plots/KT_omnes_abs_4.tex} \input{plots/KT_omnes_arg_4.tex}}
\caption{Comparison of the absolute value (left column) and phase motion (right column) of the Omn\`es function (red) and single-variable amplitude $\F_{\alpha}$ (blue) for various pion masses: first row $\mpi=0$, second row $\mpi=\bar{M}_{\pi}$, third row $\mpi=\tfrac{3}{2}\bar{M}_{\pi}$, and last row $\mpi=1.96\bar{M}_{\pi}\approx\frac{1}{3}M_{\omega}(M_\pi^2)$. The dashed black lines mark the pion-mass-dependent lower and upper phase-space boundaries for the $\wtopi$ decay given by $4\mpi^2$ and $(\mw-\mpi)^2$, respectively. The error bands are generated by taking the uncertainties of the IAM phase shift $\delta(s)$ and $\mw(\mpi^2)$ into account, see Fig.~\ref{Fig:IAMphase} and Eq.~\eqref{Eq_kin_Mw_mp}. The latter only affects $\F_{\alpha}$.}
	\label{fig:KT_omnes_abs_delta}
\end{figure*}

In this section we discuss the final results of our dispersive representation of $\wtopi$. First of all we want to compare the resulting pion-mass-dependent Omn\`es function or two-body final-state interaction\footnote{Here the third pion will act as spectator, meaning $\hat{\F}_{\alpha}(s)=0$, thus the single-variable amplitude equals the Omn\`es function in this case (see Eq.~\eqref{eq_basis_functions}).} (2-body FSI) to the single-variable amplitude $\F_{\alpha}$ (3-body FSI) to study the dynamical effects generated by the interaction with the third pion as depicted in Fig.~\ref{fig:KT_omnes_abs_delta}.

\begin{figure}
	\large
	\centering
	\scalebox{0.68}{\input{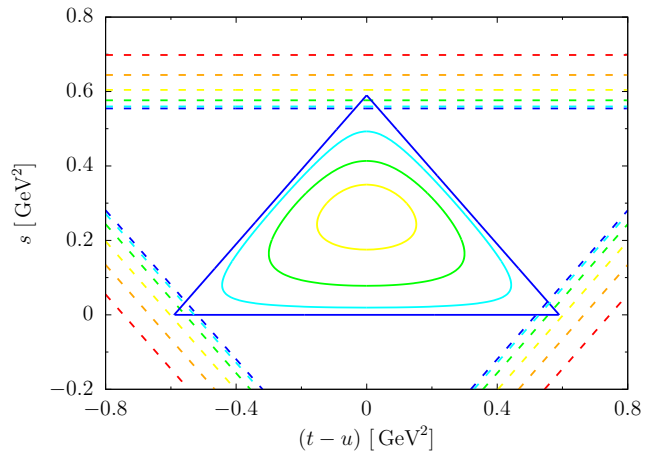}}
	\caption{Pion-mass-dependent phase space boundaries of $\wtopi$ (solid lines) and position of the ``on-shell'' $\rho$ in the respective $s$-, $t$-, and $u$-channels (dashed lines). The color coding is kept identical to Figs.~\ref{Fig:poleTrajectory} and \ref{Fig:IAMphase} with the following pion masses: 0 (blue), $\tfrac{1}{2}\bar{M}_{\pi}$ (cyan), $\bar{M}_{\pi}$ (green), $\tfrac{3}{2}\bar{M}_{\pi}$ (yellow), $2\bar{M}_{\pi}$ (orange), and $\tfrac{5}{2}\bar{M}_{\pi}$ (red). For the last two the phase space already vanishes, and thus the $\omega$ becomes stable with respect to the investigated decay mode.}
	\label{fig:phsp_mpi_dep_rho_bands}
\end{figure}

Besides pion masses close to the chiral limit, the 3-body FSI leads to an enhancement of the modulus of the single-variable amplitude compared to the Omn\`es function. While the peak position in the absolute values (due to the $\rho$ resonance) is essentially identical for 2- and 3-body FSI at a given value of $\mpi$, the phases behave rather differently. The Omn\`es function fulfills Watson's theorem and thus its phase is identical to the $\pi\pi$ scattering phase shift (the IAM phase). Due to the dynamical effects stemming from the interaction with the third pion (and the generation of a three-pion cut), this does not hold for the single-variable amplitude (see Sect.~\ref{sec:dispframework}). Thus the argument of the single-variable function is shifted compared to the input IAM phase. The lower and upper phase-space boundaries are marked by dashed vertical lines, and hence denote the kinematical range directly accessible in the decay.  Already here we want to point out that mainly the tails of the $\rho$ resonance will only contribute to the dynamics when increasing the pion mass (\cf~Fig.~\ref{fig:phsp_mpi_dep_rho_bands}).

\begin{figure}
	\large
	\centering
	\scalebox{0.68}{
		\input{plots/omega_width.tex}
	}
	\caption{Pion-mass-dependent decay width for $\wtopi$ shown for three different cases: kinematic contribution only (black), considering 2-body FSI (red), and full 3-body FSI (blue). The physical point is marked by the black diamond. The error bands are generated by taking the uncertainties of the IAM phase shift $\delta(s)$ and $\mw(\mpi^2)$ into account, see Fig.~\ref{Fig:IAMphase} and Eq.~\eqref{Eq_kin_Mw_mp}.}
	\label{fig:omega_width}
\end{figure}

In order to study the dynamical effects on the decay width we consider three different scenarios: first we consider only kinematic contributions to the decay width (all pion--pion dynamics are disregarded, $\delta(s)=0$, and thus $|\F(s,t,u)|^2=const.$), secondly we allow for 2-body rescattering effects (meaning $|\F(s,t,u)|^2 \propto |\Omega(s)+\Omega(t)+\Omega(u)|^2$), and third the full 3-body dynamics are taken into account. A comparison of the different cases is displayed in Fig.~\ref{fig:omega_width}.

\bsp
First of all we notice that the $\wtopi$ width decreases with increasing pion mass. This is not surprising since the mass of the three pions is increasing faster than the mass of the $\omega$, \cf~Eq.~\eqref{Eq_kin_Mw_mp}. Hence the phase space shrinks for larger pion masses as depicted in Fig.~\ref{fig:phsp_mpi_dep_rho_bands}. For $\mpi>\tfrac{1}{3}\mw\approx1.96\bar{M}_{\pi}$ the masses of the three pions exceed the $\omega$ mass, thus the reaction $\wtopi$ is no longer allowed and the $\omega$ becomes stable with respect to the considered decay channel (\ie, in QCD in the isospin limit).
\esp

\bsp
We now study the effects of the 2- and 3-body dynamics. First of all we notice that the kinematical prefactor given in Eq.~\eqref{eq:kin_prefactor} vanishes at the phase space boundaries in all directions. This leads to a stronger weighting of the inner region compared to the outskirts of the phase space when evaluating the integral Eq.~\eqref{Eq_DecWid_LorInvPhaseSpInt}. Secondly, the dynamics are mainly governed by the $\rho$ resonance in the respective 2-body channels, leading to a three-band structure (one for each channel) in the function $\F(s,t,u)$. The peak position and width of these bands will be determined by the respective 2- and 3-body FSI effects and the information of the scattering phase input from the IAM. We conclude that strong imprints of the $\rho$ as a dynamical effect will only affect the decay width close to the chiral limit, since the $\rho$ is only allowed to go on-shell within the phase space boundaries for $\mpi<0.15\bar{M}_{\pi}$. This effect is even reinforced by the increasing mass of the $\rho$ and its decreasing width with growing pion mass, as investigated in Sect.~\ref{sec:rho}. Thus the main contributions to the $\omega$ width will come from the tail of the $\rho$ resonances at low pion masses, see Fig.~\ref{fig:phsp_mpi_dep_rho_bands}.
\esp

\begin{figure}
	\large
	\centering
	\scalebox{0.68}{\input{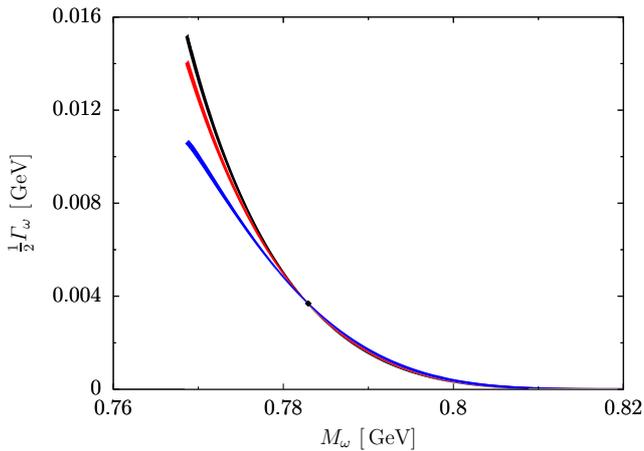}}
	\caption{Trajectory of the pion-mass-dependent $\omega$ mass and width in the complex-energy plane of $3\pi\to3\pi$ scattering for the three investigated scenarios: pure kinematics (black), 2-body FSI (red), and full 3-body FSI (blue). This plot can be compared with the one for the $\rho$ resonance, see Fig.~\ref{Fig:poleTrajectory}. The physical point is marked by the black diamond. The error bands are generated by taking the uncertainties of the IAM phase shift $\delta(s)$, $\mw(\mpi^2)$, and $\bar{\Gamma}_{\omega}$ into account, see Fig.~\ref{Fig:IAMphase}, Eq.~\eqref{Eq_kin_Mw_mp}, and Ref.~\cite{Tanabashi:2018}.}
	\label{fig:omega_pole_trajectory}
\end{figure}

\bsp
This is consistent with the result given in Fig.~\ref{fig:omega_width}. We observe that the impact of the 2- and 3-body dynamics on the decay width is very strong close to the chiral limit, where it leads to a reduction of the decay width compared to the one of pure kinematics. The damping due to the 3-body dynamics is four times stronger than the influence of the 2-body effects and results in a width smaller by about one third than the purely kinematic effects. In the chiral limit we find
\begin{align}
\Gamma_{\omega}^{\text{kinematic}}(0) &= 4.02(1)\,\bar{\Gamma}_{\omega} = 30.4(1)\MeV\,, \notag\\
\Gamma_{\omega}^{\text{2-body FSI}}(0) &= 3.72(2)\,\bar{\Gamma}_{\omega} = 28.1(1)\MeV\,, \notag\\
\Gamma_{\omega}^{\text{3-body FSI}}(0) &= 2.81(2)\,\bar{\Gamma}_{\omega} = 21.3(2)\MeV\,,
\end{align}
for the widths of the three discussed scenarios, where we took $\bar{\Gamma}_{\omega}\equiv\bar{\Gamma}(\wtopi)=7.57(9)\MeV$~\cite{Tanabashi:2018}. Since the subtraction constant of all three curves is fixed at the physical point, the difference between them shrinks when approaching this point. Above the behavior is opposite, here the 2- and 3-body dynamics generate a decay width that is larger than the one obtained from pure kinematic effects. Since the phase space at the physical point and above does not allow for on-shell $\rho$ mesons in the respective 2-body systems, the influence of the dynamical effects above this point is very small. Thus the 2- and 3-body FSI curves do not differ strongly from the curve of pure kinematics. 

Considering Eq.~\eqref{spole_M_Gamma_rho} we are in the position to predict a pion-mass-dependent trajectory of the $\omega$ pole position in the complex-energy plane of $3\pi\to3\pi$ scattering similar to the $\rho$ resonance in Fig.~\ref{Fig:poleTrajectory}. A plot of this pole trajectory is given in Fig.~\ref{fig:omega_pole_trajectory}. Note that, strictly speaking, our formalism does not determine such a complex pole position: rather, we have discussed mass (via effective Lagrangians) and width (via the Khuri--Treiman formalism) individually. Given the smallness of the width of the $\omega$, we regard the error committed thereby as negligible, although this may not obviously be so in the general case of an arbitrary $3\pi$ resonance.
\esp

\begin{figure}
	\large
	\centering
	\scalebox{0.68}{
		\input{plots/phi_width.tex}
	}
	\caption{Pion-mass-dependent decay width for $\phi\to3\pi$ shown for three different cases: kinematic contribution only (black), considering 2-body FSI (red) and full 3-body FSI (blue). The physical point is marked by the black diamond. The error bands are generated by taking the uncertainty of the IAM phase shift $\delta(s)$ into account, see Fig.~\ref{Fig:IAMphase}.}
	\label{fig:phi_width}
\end{figure}

Since dynamical effects in the investigated process are limited due to the small phase space, we want to emphasize that the dispersive representation derived in Sect.~\ref{sec:dispframework} is valid for general $V\to3\pi$ decays, with $V$ denoting an arbitrary isoscalar vector meson, at least to the extent that the elastic approximation in the pion--pion rescattering is justifiable. Thus we are able to describe $\phi\to3\pi$ by just replacing the decay mass according to $\mw\mapsto M_{\phi}$ within the same formalism~\cite{Niecknig:2012sj,Danilkin:2014cra}. With a mass of $1019\MeV$~\cite{Tanabashi:2018} its decay allows for much richer dynamics due to the larger phase space, in particular the three $\rho$ bands are visible inside the Dalitz plot~\cite{Aloisio:2003ur,Akhmetshin:2006sc}. In the quark-model picture, the $\phi$ can be understood as a pure $s\bar{s}$ state, thus its mass does not depend on the pion mass at leading order. This follows from Eq.~\eqref{Eq:mV_mpi_dependence} when using the GMOR relation to relate $M_{K}^{2}=B(\hat{m}+m_{s})+\Order(m_{q}^{2})$. The results for the pion-mass-dependent partial decay width $\Gamma(\phi\to3\pi)$ are displayed in Fig.~\ref{fig:phi_width}. Since the $\rho$ is allowed to go on-shell up to $\mpi\approx1.7\bar{M}_{\pi}$, the dynamical imprints are much stronger compared to the case of the $\wtopi$ even above the physical point.

Overall, given how subtle rescattering effects beyond two-body rescattering are usually thought to be (and more often than not neglected altogether in experimental Dalitz plot analyses), it is remarkable to see that in both cases studied here, $\wtopi$ and $\phi\to3\pi$, the 3-body FSI effects tend to affect the quark-mass dependence of the (partial) widths about as strongly as the 2-body FSI.

We wish to add a few caveats concerning the precision of the predictions shown in this section.  
Error bands are estimated solely based on the uncertainty within the one-loop IAM phase shift input, as well
as the one in $M_\omega(M_\pi^2)$, which is nonetheless based on the large-$N_c$ expansion and leading-order 
symmetry breaking in the quark masses.   We do not attempt to quantify corrections due to higher orders 
in either case, and refer to Ref.~\cite{Niehus:2018} for future work concerning an improved pion-mass-dependent
phase shift.  Furthermore, a high-precision measurement of the $\phi\to3\pi$ Dalitz plot~\cite{Aloisio:2003ur} 
revealed the need to include a \textit{second} subtraction in the dispersive representation of the decay 
amplitude~\cite{Niecknig:2012sj}, which has also not been considered here; experimental data on the $\wtopi$
Dalitz plot is not conclusive in this respect yet~\cite{Adlarson:2016wkw}.

%-----------------------------------------------------------------------------------------
\section{Summary}\label{sec:summary}
%-----------------------------------------------------------------------------------------
\bsp
We have investigated the pion-mass (and hence the quark-mass) dependence of the $\wtopi$ decay width,
generalizing previous studies of two-pion resonances based on the inverse amplitude method.
To this end, we have employed a dispersive formalism, based on Khuri--Treiman equations,
that uses inverse-amplitude-method phase shifts as input.  The pion-mass dependence of the $\omega$ mass is estimated
using a symmetry relation based on chiral perturbation theory for vector mesons.  
Deviations from phase space behavior alone, induced by the pion-mass-dependent decay amplitude, are clearly 
visible, although suppressed for larger-than-physical pion masses due to the smallness of phase space.
We have demonstrated for the decay of the heavier $\phi$ into three pions that this need not be the case
in general.  Remarkably, a simple description of the decays in terms of two-body rescattering alone does
not yield a good approximation to the full pion-mass dependence seen.

The three-pion decays of the lightest isoscalar vector mesons only serve as a paradigm case for the investigation
of three-body resonances; extensions to other, similar decays within the same formalism ought to be tested in the future.  
\esp

%-----------------------------------------------------------------------------------------
\begin{acknowledgements}
\bsp
We thank C.~Hanhart, M.~Hoferichter, M.~Niehus, J.~Ruiz de Elvira, M.~Ueding, and M.~Werner for useful discussions,
as well as M.~Hoferichter, S.~Holz, and M.~Niehus for carefully reading this manuscript.
Partial financial support by DFG and NSFC through funds provided to
the Sino--German CRC~110 \enquote{Symmetries and the Emergence of
Structure in QCD} (DFG Grant No.~TRR110 and NSFC Grant No.~11621131001) 
and by the Bonn--Cologne Graduate School of Physics and Astronomy is gratefully acknowledged.
\esp
\end{acknowledgements}
%-----------------------------------------------------------------------------------------

%-----------------------------------------------------------------------------------------
\bibliographystyle{utphysmod}
\bibliography{Literature}

\providecommand{\href}[2]{#2}\begingroup\raggedright\begin{thebibliography}{10}

\bibitem{Aoki:2016frl}
S.~Aoki {\em et~al.} [FLAG Collaboration], Eur. Phys. J. C {\bfseries 77}, 112
  (2017)
[\href{https://arxiv.org/abs/1607.00299}{{arXiv:1607.00299 [hep-lat]}}].
%%CITATION = ARXIV:1607.00299;%%.

\bibitem{Briceno:2017max}
R.~A. Brice{\~n}o, J.~J. Dudek, and R.~D. Young, Rev. Mod. Phys. {\bfseries
  90}, 025001 (2018)
[\href{https://arxiv.org/abs/1706.06223}{{arXiv:1706.06223 [hep-lat]}}].
%%CITATION = ARXIV:1706.06223;%%.

\bibitem{Weinberg:1978kz}
S.~Weinberg,
Physica A {\bfseries 96}, 327 (1979).
%%CITATION = PHYSA,A96,327;%%.

\bibitem{Gasser:1983yg}
J.~Gasser and H.~Leutwyler,
Annals Phys. {\bfseries 158}, 142 (1984).
%%CITATION = APNYA,158,142;%%.

\bibitem{Gasser:1984gg}
J.~Gasser and H.~Leutwyler,
Nucl. Phys. B {\bfseries 250}, 465 (1985).
%%CITATION = NUPHA,B250,465;%%.

\bibitem{Truong:1991gv}
T.~N. Truong,
Phys. Rev. Lett. {\bfseries 67}, 2260 (1991).
%%CITATION = PRLTA,67,2260;%%.

\bibitem{Dobado:1992ha}
A.~Dobado and J.~R. Pel{\'a}ez, Phys. Rev. D {\bfseries 47}, 4883 (1993)
[\href{https://arxiv.org/abs/hep-ph/9301276}{{arXiv:hep-ph/9301276}}].
%%CITATION = HEP-PH/9301276;%%.

\bibitem{Dobado:1996ps}
A.~Dobado and J.~R. Pel{\'a}ez, Phys. Rev. D {\bfseries 56}, 3057 (1997)
[\href{https://arxiv.org/abs/hep-ph/9604416}{{arXiv:hep-ph/9604416}}].
%%CITATION = HEP-PH/9604416;%%.

\bibitem{GomezNicola:2007qj}
A.~G{\'o}mez~Nicola, J.~R. Pel{\'a}ez, and G.~R{\'i}os, Phys. Rev. D {\bfseries
  77}, 056006 (2008)
[\href{https://arxiv.org/abs/0712.2763}{{arXiv:0712.2763 [hep-ph]}}].
%%CITATION = ARXIV:0712.2763;%%.

\bibitem{Hanhart:2008mx}
C.~Hanhart, J.~R. Pel{\'a}ez, and G.~R{\'i}os, Phys. Rev. Lett. {\bfseries
  100}, 152001 (2008)
[\href{https://arxiv.org/abs/0801.2871}{{arXiv:0801.2871 [hep-ph]}}].
%%CITATION = ARXIV:0801.2871;%%.

\bibitem{Nebreda:2010wv}
J.~Nebreda and J.~R. Pel{\'a}ez., Phys. Rev. D {\bfseries 81}, 054035 (2010)
[\href{https://arxiv.org/abs/1001.5237}{{arXiv:1001.5237 [hep-ph]}}].
%%CITATION = ARXIV:1001.5237;%%.

\bibitem{Polejaeva:2012ut}
K.~Polejaeva and A.~Rusetsky, Eur. Phys. J. A {\bfseries 48}, 67 (2012)
[\href{https://arxiv.org/abs/1203.1241}{{arXiv:1203.1241 [hep-lat]}}].
%%CITATION = ARXIV:1203.1241;%%.

\bibitem{Briceno:2012rv}
R.~A. Brice{\~n}o and Z.~Davoudi, Phys. Rev. D {\bfseries 87}, 094507 (2013)
[\href{https://arxiv.org/abs/1212.3398}{{arXiv:1212.3398 [hep-lat]}}].
%%CITATION = ARXIV:1212.3398;%%.

\bibitem{Hansen:2014eka}
M.~T. Hansen and S.~R. Sharpe, Phys. Rev. D {\bfseries 90}, 116003 (2014)
[\href{https://arxiv.org/abs/1408.5933}{{arXiv:1408.5933 [hep-lat]}}].
%%CITATION = ARXIV:1408.5933;%%.

\bibitem{Hansen:2015zga}
M.~T. Hansen and S.~R. Sharpe, Phys. Rev. D {\bfseries 92}, 114509 (2015)
[\href{https://arxiv.org/abs/1504.04248}{{arXiv:1504.04248 [hep-lat]}}].
%%CITATION = ARXIV:1504.04248;%%.

\bibitem{Hammer:2017uqm}
H.-W. Hammer, J.-Y. Pang, and A.~Rusetsky, JHEP {\bfseries 1709}, 109 (2017)
[\href{https://arxiv.org/abs/1706.07700}{{arXiv:1706.07700 [hep-lat]}}].
%%CITATION = ARXIV:1706.07700;%%.

\bibitem{Hammer:2017kms}
H.-W. Hammer, J.-Y. Pang, and A.~Rusetsky, JHEP {\bfseries 1710}, 115 (2017)
[\href{https://arxiv.org/abs/1707.02176}{{arXiv:1707.02176 [hep-lat]}}].
%%CITATION = ARXIV:1707.02176;%%.

\bibitem{Mai:2017bge}
M.~Mai and M.~D{\"o}ring, Eur. Phys. J. A {\bfseries 53}, 240 (2017)
[\href{https://arxiv.org/abs/1709.08222}{{arXiv:1709.08222 [hep-lat]}}].
%%CITATION = ARXIV:1709.08222;%%.

\bibitem{Khuri:1960zz}
N.~N. Khuri and S.~B. Treiman,
Phys. Rev. {\bfseries 119}, 1115 (1960).
%%CITATION = PHRVA,119,1115;%%.

\bibitem{Guo:2008nc}
F.-K. Guo, C.~Hanhart, F.~J. Llanes-Estrada, and U.-G. Mei{\ss}ner, Phys. Lett.
  B {\bfseries 678}, 90 (2009)
[\href{https://arxiv.org/abs/0812.3270}{{arXiv:0812.3270 [hep-ph]}}].
%%CITATION = ARXIV:0812.3270;%%.

\bibitem{Niehus:2017}
M.~Niehus, Master's thesis, University of Bonn  (2017).

\bibitem{Pelaez:2010fj}
J.~R. Pel{\'a}ez and G.~R{\'i}os, Phys. Rev. D {\bfseries 82}, 114002 (2010)
[\href{https://arxiv.org/abs/1010.6008}{{arXiv:1010.6008 [hep-ph]}}].
%%CITATION = ARXIV:1010.6008;%%.

\bibitem{GellMann:1968rz}
M.~Gell-Mann, R.~J. Oakes, and B.~Renner,
Phys. Rev. {\bfseries 175}, 2195 (1968).
%%CITATION = PHRVA,175,2195;%%.

\bibitem{Bazavov:2010hj}
A.~Bazavov {\em et~al.} [MILC Collaboration], PoS LATTICE {\bfseries 2010}, 074
  (2010)
[\href{https://arxiv.org/abs/1012.0868}{{arXiv:1012.0868 [hep-lat]}}].
%%CITATION = ARXIV:1012.0868;%%.

\bibitem{Beane:2011zm}
S.~R. Beane {\em et~al.}, Phys. Rev. D {\bfseries 86}, 094509 (2012)
[\href{https://arxiv.org/abs/1108.1380}{{arXiv:1108.1380 [hep-lat]}}].
%%CITATION = ARXIV:1108.1380;%%.

\bibitem{Borsanyi:2012zv}
S.~Bors{\'a}nyi, S.~D{\"u}rr, Z.~Fodor, S.~Krieg, A.~Sch{\"a}fer, E.~E. Scholz,
  and K.~K. Szab{\'o}, Phys. Rev. D {\bfseries 88}, 014513 (2013)
[\href{https://arxiv.org/abs/1205.0788}{{arXiv:1205.0788 [hep-lat]}}].
%%CITATION = ARXIV:1205.0788;%%.

\bibitem{Durr:2013goa}
S.~D{\"u}rr {\em et~al.} [BMW Collaboration], Phys. Rev. D {\bfseries 90},
  114504 (2014)
[\href{https://arxiv.org/abs/1310.3626}{{arXiv:1310.3626 [hep-lat]}}].
%%CITATION = ARXIV:1310.3626;%%.

\bibitem{Blum:2014tka}
T.~Blum {\em et~al.} [RBC, UKQCD Collaboration], Phys. Rev. D {\bfseries 93},
  074505 (2016)
[\href{https://arxiv.org/abs/1411.7017}{{arXiv:1411.7017 [hep-lat]}}].
%%CITATION = ARXIV:1411.7017;%%.

\bibitem{Tanabashi:2018}
M.~Tanabashi {\em et~al.} [Particle Data Group], Phys. Rev. D {\bfseries 98},
  030001 (2018).

\bibitem{Gribov:2009zz}
V.~N. Gribov, {\em {Strong interactions of hadrons at high energies: Gribov
  lectures on theoretical physics}}, Cambridge University Press,
2012.
\newblock
%%CITATION = CMPCE,27,;%%.

\bibitem{Moussallam:2011zg}
B.~Moussallam, Eur. Phys. J. C {\bfseries 71}, 1814 (2011)
[\href{https://arxiv.org/abs/1110.6074}{{arXiv:1110.6074 [hep-ph]}}].
%%CITATION = ARXIV:1110.6074;%%.

\bibitem{Colangelo:2001df}
G.~Colangelo, J.~Gasser, and H.~Leutwyler, Nucl. Phys. B {\bfseries 603}, 125
  (2001)
[\href{https://arxiv.org/abs/hep-ph/0103088}{{arXiv:hep-ph/0103088}}].
%%CITATION = HEP-PH/0103088;%%.

\bibitem{GarciaMartin:2011jx}
R.~Garc{\'i}a-Mart{\'i}n, R.~Kami{\'n}ski, J.~R. Pel{\'a}ez, and J.~Ruiz~de
  Elvira, Phys. Rev. Lett. {\bfseries 107}, 072001 (2011)
[\href{https://arxiv.org/abs/1107.1635}{{arXiv:1107.1635 [hep-ph]}}].
%%CITATION = ARXIV:1107.1635;%%.

\bibitem{Niehus:2018}
M.~Niehus {\em et~al.}, work in progress.

\bibitem{Bruns:2017gix}
P.~C. Bruns and M.~Mai, Phys. Lett. B {\bfseries 778}, 43 (2018)
[\href{https://arxiv.org/abs/1707.08983}{{arXiv:1707.08983 [hep-lat]}}].
%%CITATION = ARXIV:1707.08983;%%.

\bibitem{Alexandrou:2017mpi}
C.~Alexandrou {\em et~al.}, Phys. Rev. D {\bfseries 96}, 034525 (2017)
[\href{https://arxiv.org/abs/1704.05439}{{arXiv:1704.05439 [hep-lat]}}].
%%CITATION = ARXIV:1704.05439;%%.

\bibitem{Aoki:2007rd}
S.~Aoki {\em et~al.} [CP-PACS Collaboration], Phys. Rev. D {\bfseries 76},
  094506 (2007)
[\href{https://arxiv.org/abs/0708.3705}{{arXiv:0708.3705 [hep-lat]}}].
%%CITATION = ARXIV:0708.3705;%%.

\bibitem{Feng:2010es}
X.~Feng, K.~Jansen, and D.~B. Renner, Phys. Rev. D {\bfseries 83}, 094505
  (2011)
[\href{https://arxiv.org/abs/1011.5288}{{arXiv:1011.5288 [hep-lat]}}].
%%CITATION = ARXIV:1011.5288;%%.

\bibitem{Aoki:2011yj}
S.~Aoki {\em et~al.} [CS Collaboration], Phys. Rev. D {\bfseries 84}, 094505
  (2011)
[\href{https://arxiv.org/abs/1106.5365}{{arXiv:1106.5365 [hep-lat]}}].
%%CITATION = ARXIV:1106.5365;%%.

\bibitem{Lang:2011mn}
C.~B. Lang, D.~Mohler, S.~Prelovsek, and M.~Vidmar, Phys. Rev. D {\bfseries
  84}, 054503 (2011) [\href{https://arxiv.org/abs/1105.5636}{{arXiv:1105.5636
  [hep-lat]}}]
  [Erratum: Phys. Rev. D {\bf 89}, 059903 (2014)].
%%CITATION = ARXIV:1105.5636;%%.

\bibitem{Dudek:2012xn}
J.~J. Dudek, R.~G. Edwards, and C.~E. Thomas [Hadron Spectrum Collaboration],
  Phys. Rev. D {\bfseries 87}, 034505 (2013)
  [\href{https://arxiv.org/abs/1212.0830}{{arXiv:1212.0830 [hep-ph]}}]
  [Erratum: Phys. Rev. D {\bf 90}, 099902 (2014)].
%%CITATION = ARXIV:1212.0830;%%.

\bibitem{Pelissier:2012pi}
C.~Pelissier and A.~Alexandru, Phys. Rev. D {\bfseries 87}, 014503 (2013)
[\href{https://arxiv.org/abs/1211.0092}{{arXiv:1211.0092 [hep-lat]}}].
%%CITATION = ARXIV:1211.0092;%%.

\bibitem{Bali:2015gji}
G.~S. Bali {\em et~al.} [RQCD Collaboration], Phys. Rev. D {\bfseries 93},
  054509 (2016)
[\href{https://arxiv.org/abs/1512.08678}{{arXiv:1512.08678 [hep-lat]}}].
%%CITATION = ARXIV:1512.08678;%%.

\bibitem{Wilson:2015dqa}
D.~J. Wilson, R.~A. Brice{\~n}o, J.~J. Dudek, R.~G. Edwards, and C.~E. Thomas,
  Phys. Rev. D {\bfseries 92}, 094502 (2015)
[\href{https://arxiv.org/abs/1507.02599}{{arXiv:1507.02599 [hep-ph]}}].
%%CITATION = ARXIV:1507.02599;%%.

\bibitem{Bulava:2016mks}
J.~Bulava, B.~Fahy, B.~H{\"o}rz, K.~J. Juge, C.~Morningstar, and C.~H. Wong,
  Nucl. Phys. B {\bfseries 910}, 842 (2016)
[\href{https://arxiv.org/abs/1604.05593}{{arXiv:1604.05593 [hep-lat]}}].
%%CITATION = ARXIV:1604.05593;%%.

\bibitem{Hu:2017wli}
B.~Hu, R.~Molina, M.~D{\"o}ring, M.~Mai, and A.~Alexandru, Phys. Rev. D
  {\bfseries 96}, 034520 (2017)
[\href{https://arxiv.org/abs/1704.06248}{{arXiv:1704.06248 [hep-lat]}}].
%%CITATION = ARXIV:1704.06248;%%.

\bibitem{Andersen:2018mau}
C.~Andersen, J.~Bulava, B.~H{\"o}rz, and C.~Morningstar,
\href{https://arxiv.org/abs/1808.05007}{{arXiv:1808.05007 [hep-lat]}}.
%%CITATION = ARXIV:1808.05007;%%.

\bibitem{Caprini:2011ky}
I.~Caprini, G.~Colangelo, and H.~Leutwyler, Eur. Phys. J. C {\bfseries 72},
  1860 (2012)
[\href{https://arxiv.org/abs/1111.7160}{{arXiv:1111.7160 [hep-ph]}}].
%%CITATION = ARXIV:1111.7160;%%.

\bibitem{GarciaMartin:2011cn}
R.~Garc{\'i}a-Mart{\'i}n, R.~Kami{\'n}ski, J.~R. Pel{\'a}ez, J.~Ruiz~de Elvira,
  and F.~J. Yndur{\'a}in, Phys. Rev. D {\bfseries 83}, 074004 (2011)
[\href{https://arxiv.org/abs/1102.2183}{{arXiv:1102.2183 [hep-ph]}}].
%%CITATION = ARXIV:1102.2183;%%.

\bibitem{Kibble:1960zz}
T.~W.~B. Kibble,
Phys. Rev. {\bfseries 117}, 1159 (1960).
%%CITATION = PHRVA,117,1159;%%.

\bibitem{Niecknig:2012sj}
F.~Niecknig, B.~Kubis, and S.~P. Schneider, Eur. Phys. J. C {\bfseries 72},
  2014 (2012)
[\href{https://arxiv.org/abs/1203.2501}{{arXiv:1203.2501 [hep-ph]}}].
%%CITATION = ARXIV:1203.2501;%%.

\bibitem{Hoferichter:2012pm}
M.~Hoferichter, B.~Kubis, and D.~Sakkas, Phys. Rev. D {\bfseries 86}, 116009
  (2012)
[\href{https://arxiv.org/abs/1210.6793}{{arXiv:1210.6793 [hep-ph]}}].
%%CITATION = ARXIV:1210.6793;%%.

\bibitem{Danilkin:2014cra}
I.~V. Danilkin, C.~Fern{\'a}ndez-Ram{\'i}rez, P.~Guo, V.~Mathieu, D.~Schott,
  M.~Shi, and A.~P. Szczepaniak, Phys. Rev. D {\bfseries 91}, 094029 (2015)
[\href{https://arxiv.org/abs/1409.7708}{{arXiv:1409.7708 [hep-ph]}}].
%%CITATION = ARXIV:1409.7708;%%.

\bibitem{Hoferichter:2014vra}
M.~Hoferichter, B.~Kubis, S.~Leupold, F.~Niecknig, and S.~P. Schneider, Eur.
  Phys. J. C {\bfseries 74}, 3180 (2014)
[\href{https://arxiv.org/abs/1410.4691}{{arXiv:1410.4691 [hep-ph]}}].
%%CITATION = ARXIV:1410.4691;%%.

\bibitem{Hoferichter:2017ftn}
M.~Hoferichter, B.~Kubis, and M.~Zanke, Phys. Rev. D {\bfseries 96}, 114016
  (2017)
[\href{https://arxiv.org/abs/1710.00824}{{arXiv:1710.00824 [hep-ph]}}].
%%CITATION = ARXIV:1710.00824;%%.

\bibitem{Hoferichter:2018kwz}
M.~Hoferichter, B.-L. Hoid, B.~Kubis, S.~Leupold, and S.~P. Schneider, JHEP
  {\bfseries 10}, 141 (2018)
[\href{https://arxiv.org/abs/1808.04823}{{arXiv:1808.04823 [hep-ph]}}].
%%CITATION = ARXIV:1808.04823;%%.

\bibitem{Stern:1993rg}
J.~Stern, H.~Sazdjian, and N.~H. Fuchs, Phys. Rev. D {\bfseries 47}, 3814
  (1993)
[\href{https://arxiv.org/abs/hep-ph/9301244}{{arXiv:hep-ph/9301244}}].
%%CITATION = HEP-PH/9301244;%%.

\bibitem{Knecht:1995tr}
M.~Knecht, B.~Moussallam, J.~Stern, and N.~H. Fuchs, Nucl. Phys. B {\bfseries
  457}, 513 (1995)
[\href{https://arxiv.org/abs/hep-ph/9507319}{{arXiv:hep-ph/9507319}}].
%%CITATION = HEP-PH/9507319;%%.

\bibitem{Ananthanarayan:2000cp}
B.~Ananthanarayan and P.~B{\"u}ttiker, Eur. Phys. J. C {\bfseries 19}, 517
  (2001)
[\href{https://arxiv.org/abs/hep-ph/0012023}{{arXiv:hep-ph/0012023}}].
%%CITATION = HEP-PH/0012023;%%.

\bibitem{Zdrahal:2008bd}
M.~Zdr{\'a}hal and J.~Novotn{\'y}, Phys. Rev. D {\bfseries 78}, 116016 (2008)
[\href{https://arxiv.org/abs/0806.4529}{{arXiv:0806.4529 [hep-ph]}}].
%%CITATION = ARXIV:0806.4529;%%.

\bibitem{Hoferichter:2018dmo}
M.~Hoferichter, B.-L. Hoid, B.~Kubis, S.~Leupold, and S.~P. Schneider, Phys.
  Rev. Lett. {\bfseries 121}, 112002 (2018)
[\href{https://arxiv.org/abs/1805.01471}{{arXiv:1805.01471 [hep-ph]}}].
%%CITATION = ARXIV:1805.01471;%%.

\bibitem{Bronzan:1963mby}
J.~B. Bronzan and C.~Kacser,
Phys. Rev. {\bfseries 132}, 2703 (1963).
%%CITATION = PHRVA,132,2703;%%.

\bibitem{Gribov:1962fu}
V.~N. Gribov, V.~V. Anisovich, and A.~A. Anselm, Sov. Phys. JETP {\bfseries
  15}, 159 (1962)
  [Zh. Eksp. Teor. Fiz. {\bf 42}, 224 (1962)].
%%CITATION = SPHJA,15,159;%%.

\bibitem{Froissart:1961ux}
M.~Froissart,
Phys. Rev. {\bfseries 123}, 1053 (1961).
%%CITATION = PHRVA,123,1053;%%.

\bibitem{Martin:1962rt}
A.~Martin,
Phys. Rev. {\bfseries 129}, 1432 (1963).
%%CITATION = PHRVA,129,1432;%%.

\bibitem{Omnes:1958hv}
R.~Omn{\`e}s,
Nuovo Cim. {\bfseries 8}, 316 (1958).
%%CITATION = NUCIA,8,316;%%.

\bibitem{Jenkins:1995vb}
E.~E. Jenkins, A.~V. Manohar, and M.~B. Wise, Phys. Rev. Lett. {\bfseries 75},
  2272 (1995)
[\href{https://arxiv.org/abs/hep-ph/9506356}{{arXiv:hep-ph/9506356}}].
%%CITATION = HEP-PH/9506356;%%.

\bibitem{Bijnens:1996nq}
J.~Bijnens and P.~Gosdzinsky, Phys. Lett. B {\bfseries 388}, 203 (1996)
[\href{https://arxiv.org/abs/hep-ph/9607462}{{arXiv:hep-ph/9607462}}].
%%CITATION = HEP-PH/9607462;%%.

\bibitem{Bijnens:1997ni}
J.~Bijnens, P.~Gosdzinsky, and P.~Talavera, Nucl. Phys. B {\bfseries 501}, 495
  (1997)
[\href{https://arxiv.org/abs/hep-ph/9704212}{{arXiv:hep-ph/9704212}}].
%%CITATION = HEP-PH/9704212;%%.

\bibitem{Pelaez:2016klv}
J.~R. Pel{\'a}ez, A.~Rodas, and J.~Ruiz~de Elvira, Eur. Phys. J. C {\bfseries
  77}, 91 (2017)
[\href{https://arxiv.org/abs/1612.07966}{{arXiv:1612.07966 [hep-ph]}}].
%%CITATION = ARXIV:1612.07966;%%.

\bibitem{Bruns:2004tj}
P.~C. Bruns and U.-G. Mei{\ss}ner, Eur. Phys. J. C {\bfseries 40}, 97 (2005)
[\href{https://arxiv.org/abs/hep-ph/0411223}{{arXiv:hep-ph/0411223}}].
%%CITATION = HEP-PH/0411223;%%.

\bibitem{Bruns:2013tja}
P.~C. Bruns, L.~Greil, and A.~Sch{\"a}fer, Phys. Rev. D {\bfseries 88}, 114503
  (2013)
[\href{https://arxiv.org/abs/1309.3976}{{arXiv:1309.3976 [hep-ph]}}].
%%CITATION = ARXIV:1309.3976;%%.

\bibitem{Bavontaweepanya:2018yds}
R.~Bavontaweepanya, X.-Y. Guo, and M.~F.~M. Lutz, Phys. Rev. D {\bfseries 98},
  056005 (2018)
[\href{https://arxiv.org/abs/1801.10522}{{arXiv:1801.10522 [hep-ph]}}].
%%CITATION = ARXIV:1801.10522;%%.

\bibitem{GellMann:1962jt}
M.~Gell-Mann, D.~Sharp, and W.~G. Wagner,
Phys. Rev. Lett. {\bfseries 8}, 261 (1962).
%%CITATION = PRLTA,8,261;%%.

\bibitem{Klingl:1996by}
F.~Klingl, N.~Kaiser, and W.~Weise, Z. Phys. A {\bfseries 356}, 193 (1996)
[\href{https://arxiv.org/abs/hep-ph/9607431}{{arXiv:hep-ph/9607431}}].
%%CITATION = HEP-PH/9607431;%%.

\bibitem{Leupold:2008bp}
S.~Leupold and M.~F.~M. Lutz, Eur. Phys. J. A {\bfseries 39}, 205 (2009)
[\href{https://arxiv.org/abs/0807.4686}{{arXiv:0807.4686 [hep-ph]}}].
%%CITATION = ARXIV:0807.4686;%%.

\bibitem{Dax:2017}
M.~Dax, Bachelor's thesis, University of Bonn  (2017).

\bibitem{Gasser:1984pr}
J.~Gasser and H.~Leutwyler,
Nucl. Phys. B {\bfseries 250}, 539 (1985).
%%CITATION = NUPHA,B250,539;%%.

\bibitem{Ditsche:2008cq}
C.~Ditsche, B.~Kubis, and U.-G. Mei{\ss}ner, Eur. Phys. J. C {\bfseries 60}, 83
  (2009)
[\href{https://arxiv.org/abs/0812.0344}{{arXiv:0812.0344 [hep-ph]}}].
%%CITATION = ARXIV:0812.0344;%%.

\bibitem{Schneider:2010hs}
S.~P. Schneider, B.~Kubis, and C.~Ditsche, JHEP {\bfseries 1102}, 028 (2011)
[\href{https://arxiv.org/abs/1010.3946}{{arXiv:1010.3946 [hep-ph]}}].
%%CITATION = ARXIV:1010.3946;%%.

\bibitem{Colangelo:2018jxw}
G.~Colangelo, S.~Lanz, H.~Leutwyler, and E.~Passemar,
\href{https://arxiv.org/abs/1807.11937}{{arXiv:1807.11937 [hep-ph]}}.
%%CITATION = ARXIV:1807.11937;%%.

\bibitem{Aloisio:2003ur}
A.~Aloisio {\em et~al.} [KLOE Collaboration], Phys. Lett. B {\bfseries 561}, 55
  (2003) [\href{https://arxiv.org/abs/hep-ex/0303016}{{arXiv:hep-ex/0303016}}]
  [Erratum: Phys. Lett. B {\bf 609}, 449 (2005)].
%%CITATION = HEP-EX/0303016;%%.

\bibitem{Akhmetshin:2006sc}
R.~R. Akhmetshin {\em et~al.},
Phys. Lett. B {\bfseries 642}, 203 (2006).
%%CITATION = PHLTA,B642,203;%%.

\bibitem{Adlarson:2016wkw}
P.~Adlarson {\em et~al.} [WASA-at-COSY Collaboration], Phys. Lett. B {\bfseries
  770}, 418 (2017)
[\href{https://arxiv.org/abs/1610.02187}{{arXiv:1610.02187 [nucl-ex]}}].\\
%%CITATION = ARXIV:1610.02187;%%.

\end{thebibliography}\endgroup

\end{document}